\documentstyle[12pt,prc,aps]{revtex}

\def\Pom{{\bf I\!P}}

\def\lsim{\mathrel{\rlap{\lower4pt\hbox{\hskip1pt$\sim$}}
    \raise1pt\hbox{$<$}}}         
\def\gsim{\mathrel{\rlap{\lower4pt\hbox{\hskip1pt$\sim$}}
    \raise1pt\hbox{$>$}}}         

\def\beq{\begin{equation}}
\def\endeq{\end{equation}}
\def\arr{\begin{eqnarray}}
\def\endarr{\end{eqnarray}}
\makeindex

\begin{document}
\draft

\title {Leading proton spectrum from DIS at HERA}

\author{A. Szczurek}

\address{Institute of Nuclear Physics, PL-31-342 Cracow, Poland}

\author{N.N. Nikolaev$^{a)}$ and J. Speth}

\address{Institut f\"ur Kernphysik, Forschungszentrum J\"ulich,D-52425
J\"ulich
Germany\\
$^{a)}$ L.D.Landau Institute for Theoretical Physics, 117334 Moscow, Russia}

%

\maketitle

\vskip 2cm

\begin{abstract}
The QCD hardness scale for secondary particles ($h$) production in
semi-inclusive deep inelastic scattering (DIS), $e p \rightarrow e' X h$,
 gradually decreases
from  $Q^{2}$, the photon virtuality which determines the hard scale in the
virtual photon
(current) fragmentation region to a soft, hadronic, scale in the proton
fragmentation region. This suggests similarity of the inclusive
spectra of leading protons and neutrons, $h=p,n$, in high energy
hadron-proton and virtual photon-proton collisions. We explore this
similarity extending to the DIS regime the nonperturbative peripheral
 mechanisms of inelastic scattering traditionally used in hadronic interactions
to explain  fast nucleons production. While the
production of leading neutrons is known to be  exhausted by DIS off
charged pions, the production of leading
protons by DIS off neutral pions must be supplemented by a
substantial contribution
from isoscalar reggeon ($f_{0}$) exchange extrapolated down to moderate
values of $x_{L}$. We comment on the $x$ and $Q^{2}$ dependence of leading
proton production as a probe of a universal pattern of the $x,Q^{2}$
evolution of the nucleon and meson (reggeon) structure functions at small $x$.
\end{abstract}

\pagebreak
\vskip 2cm
In deep inelastic $ep$ scattering, according to the standard QCD
description of hadronization, the proton fragmentation region
is very different from the current and/or photon fragmentation
region. Namely, the virtuality of partons in generalized ladder diagrams
gradually decreases from the hard scale $Q^{2}$ of the struck parton
in the current fragmentation region to the soft, hadronic, scale of the
parent parton in the proton fragmentation region.

Although until quite recently
experimental data on proton fragmentation were scarce \footnote{Before
the ZEUS/H1 LPS era, leading protons in DIS have been studied only in the
fixed-target bubble-chamber neutrino experiments \cite{BEBC}. Only large
values of the Bjorken $x$ were accessible, leading to a strong kinematical
bias in the leading proton spectra \cite{MNT}. The small-$x$ data from
HERA are free of this kinematical bias.}, presently the ZEUS and H1
leading proton spectrometers (LPS) and forward neutron calorimeters (FNC)
are operational and are amassing  data on leading proton and
neutron production \cite{ZEUS_LPS97,H1_LPS97}.

Whereas popular Monte-Carlo
implementations of perturbative QCD (Ariadne \cite{Ariadne},
Herwig \cite{Herwig} and others)
are very successful in the photon fragmentation region (for a recent review
see \cite{DeWolf}) , a purely perturbative description of the proton
fragmentation region is not yet possible and the current versions
of Monte Carlo hadronization models
underestimate the yield of fast secondary nucleons
\cite{ZEUSneutron}. Traditionally, leading proton production in inelastic
collisions is modeled via nonperturbative peripheral interactions.
Such peripheral models
were quite successful when applied to hadronic collisions (for a review
see \cite{Kaidalov}). A well known example of this type of
processes  is the interaction of projectiles
with pions from the chiral mesonic substructure of the proton.
We recall that in hadronic reactions,
the pion exchange mechanism  with absorption exhausts
the cross section for the
leading neutron production in
$pp \rightarrow  Xn$, $\pi n \rightarrow Xp$, $pn \rightarrow X n$
(see \cite{NSZ97} and references therein).
There is at present a mounting evidence of the
importance of the chiral mesonic
substructure of the nucleon also in DIS (see \cite{ST}
and the reference therein).
Here let us mention a successful explanation of the Gottfried Sum Rule
violation (see for instance \cite{HSS96} and references therein
\footnote{For related work on electro-magnetic properties of nucleons
and $W$-boson and
jet production in nucleon-nucleon collisions see\cite{DHSS97,MC_he}.}
) and of the $\bar d$-$\bar u$ asymmetry as seen in $pp$ and $pn$ Drell-Yan
production \cite{SEHS96}. Exactly the same physics opens an exciting
possibility to study the pion structure function at HERA. Here one can
separate DIS off nearly on-mass-shell $\pi^{+}$ by triggering on fast
neutrons from semi-inclusive $ep \rightarrow e'Xn$ reaction \cite{HLNSS94}
(see also \cite{Lusignoli,KPP96,PSI97}). In this communication we explore
to which extent leading proton production in DIS can be understood
quantitatively within peripheral mechanisms.

Following the conventions for diffractive
DIS \cite{H1_reggeon97}, we define the semi-inclusive structure function
\begin{equation}
\frac{ d\sigma(ep \rightarrow e'p'X) } { dx dQ^{2} dz dt } =
\frac{2 \pi\alpha^2}{Q^2 x} [ 2 - 2y + \frac{y^2}{1+R} ]
\cdot F_{2}^{(4)}(z,t,\beta,Q^2) \; .
\label{cs_semiinclusive}
\end{equation}
Here $\alpha$ is the electro-magnetic fine structure constant, $z$ is the
fraction of the light-cone momentum of the beam proton carried by the
outgoing proton $z =\frac{p'^{+}}{p^{+}}$
(the same quantity is often denoted by $x_{L}$),
$t$ is the $(p,p')$ four-momentum transfer squared,
$x$, $y$, $Q^{2}$ and $\beta={Q^{2} \over M^{2}+Q^{2}}=
{x\over 1-z}$ are the standard DIS variables and $R=\sigma_{L}/\sigma_{T}$.
In the present analysis we focus on leading protons with
$0.6 < z < 0.9$. In our extension of the  peripheral
models of fragmentation  used in
hadronic reactions to lepton DIS at HERA, we consider  four mechanisms
of leading proton production  (Fig.~1): a) diffractive production of
protons (pomeron $\Pom$ exchange) which dominates at $z \rightarrow 1$, and
 constitutes a background to fragmentation in non-diffractive DIS at
$z\lsim 0.9-0.95$; b) spectator protons from the fragmentation of
the $\pi N$ Fock state of the physical proton produced by
DIS off virtual $\pi^{0}$
(pion-exchange mechanism);
c) protons produced as decay products of fast baryon resonances of which
the $\Delta$ production via pion-exchange is a typical, and predominant,
source; d) reggeized heavy meson (reggeon $R$) exchange contribution
(predominantly the isoscalar reggeon, $R=f_{0}$, exchange).
A preliminary evaluation of the first two  mechanisms has been done in
\cite{HLNSS94}.
The background from pion and reggeon exchange to the
dominant pomeron exchange
at $z \gsim 0.95$ has been discussed recently in
\cite{H1_reggeon97,NSZ96,GK96,GKS97}, and we partially use the
results of these works. Diffraction excitation of the proton into high-mass
states also contributes to  leading proton production, and we shall
comment on this small contribution following the considerations in
\cite{HNSSZ96}.

Under approximations to be specified below, the
contributions of all four mechanisms to the semi-inclusive structure
function can be written in the factorized form ($i=\Pom,\pi^{0}p,
\pi\Delta,f_{0}$):
\begin{equation}
F_{2}^{(4)}(z,t,\beta,Q^2) =
\sum_{i} F_{2}^{(4)}(i;z,t,\beta,Q^2) =
\sum_{i} f_{i}(z,t) \cdot F_2^{i}(\beta,Q^2) \; ,
\label{Expansion}
\end{equation}
where $F_2^{i}(\beta,Q^2)$ is the structure function of the
exchanged object (pion, pomeron, reggeon), $f_{i}(z,t)$ is
its flux factor  and $\beta$ is
the Bjorken variable for DIS off the exchanged object.

We start our discussion with the pion exchange mechanism. In this case
$F_2^{i}(\beta,Q^2)$ is the structure function of the physical pion and the
flux factor is given by
\begin{equation}
f_{\pi^{0} p}(x_{\pi},t) = \frac{ g_{p\pi^{0}p}^2}{16 \pi^2}(1-z)
\frac{(-t) |F_{\pi N}(z,t)|^2}{(t - m_{\pi}^2)^2} \; .
\label{flux_piN}
\end{equation}
Strictly speaking, Eq.~(\ref{flux_piN}) holds
in the plane wave impulse approximation.
A recent analysis \cite{NSZ97} has shown that absorption corrections to
 pion exchange in DIS are small and can be neglected for the purposes of
the present analysis. Also, the off-mass shell extrapolation effects are
marginal and the on-mass shell pion structure function can be used.
Important consistency check is provided by the simultaneous description
of the hadronic leading nucleon data. The results for DIS in the
interesting region of $0.6\lsim z\lsim 0.9$ only marginally depend on
whether the light-cone or Regge parameterization of
$|F_{\pi N}(z,t)|^{2}$ are used (for a detailed discussion
concerning the choice of the form factor see Refs.\cite{NSZ97,HSS96}).

Production of fast $\Delta$'s is also known to be dominated by
pion exchange. For  $\Delta^{++}$ production the flux factor is given by
\begin{equation}
f_{\pi \Delta}(z,t) =
\frac{2 g_{p \pi^{-} \Delta^{++}}} {16 \pi^{2}}
(1-z)
\frac{((m_{\Delta} + m_{N})^2 - t)^2((m_{\Delta} - m_{N})^2 - t)
|F_{\pi \Delta}(z,t)|^2}
{6 m_{N}^2 m_{\Delta}^2 (t - m_{\pi})^2} \; .
\end{equation}
Contributions from  $\Delta^{+}$ and $\Delta^{0}$ production
can be included using the familiar isospin relations (see for instance
\cite{HSS96,HNSSZ96}). In the
simplest one-pion exchange approximation, the polarization state of the
produced $\Delta$'s is such that the $\Delta \rightarrow \pi N$ the decay
angular distribution in the Gottfried-Jackson (t-channel) frame equals:
$$
w(\theta_{J},\phi_{TY}) = 1/4 \cdot (1+3cos^2\theta_{J}) \cdot
Y_{00}(\phi_{TY}) \; ,
$$
where $\theta_J$ and and $\phi_{TY}$ are
the so-called Jackson and Treiman-Yang angles, respectively. Absorptive
correction modify slightly this simple form \cite{ZakSerg},
but these corrections
can be neglected for the purposes of the present analysis, since
both the $z-$ and $t-$ spectra of decay protons only weakly depend
on the decay angular distributions.

For the diffractive $e+p \rightarrow e'+p'+X$ reaction, our semi-inclusive
structure function coincides with the pomeron component of the diffractive
structure function,
$F_{2}^{(4)}(\Pom N;z,t,\beta,Q^2)=F_{2,\Pom}^{D(4)}(z,t,\beta,Q^2)$.
At $z\lsim 0.9$, diffractive
DIS  is a small background to  non-diffractive DIS
and a somewhat simplified description  is justified.
Since  the ZEUS data have \cite{ZEUS_LPS97} $x\lsim 10^{-3}$ and $z\lsim 0.9$
then  $\beta$ is quite small, $\beta \lsim 2\cdot 10^{-3}-10^{-2}$ and it has
been argued \cite{NZ94,GNZ95} that at such a small $\beta$ one expects
the factorization
\begin{equation}
F_{2}^{D(4)}(z,t,\beta,Q^2)= f_{\Pom}(z,t)
\cdot F_{2}^{\Pom}(\beta,Q^{2}) \; .
\label{F2dif}
\end{equation}
The normalization of the pomeron flux factor $f_{\Pom}(z,t)$ and the
pomeron structure function $F_{2}^{\Pom}(\beta,Q^{2})$ is a matter
of convention, and only the product of the two is well defined. To be
specific, we use the triple-Regge parameterization for the flux factor
\begin{equation}
f_{\Pom}(z,t)= {1 \over 8 \pi^{2}(1-z)}
(1-z)^{2(1-\alpha_{\Pom}(t))} G_{\Pom}(t) ,
\label{Pomflux}
\end{equation}
where $G_{\Pom}(t) = G_{0} \exp(B_{\Pom}t)$ with $G_{0}= 21.2$ mb
\cite{NSZ96,GK96,GKS97} from the Regge decomposition of the NN total cross
sections \cite{DLsigma} and $B_{\Pom} = 3.8  $GeV$^{-2}$ according to the
triple-Regge analysis of hadronic diffraction scattering
\cite{Kaidalov,FFox,KKLP76,DLdif}. For $z \lsim 0.9$,
the specific Regge effects coming from $(1-z)^{2(1-\alpha_{\Pom}(t))}$ and
from the $t$-dependence of the pomeron trajectory $\alpha_{\Pom}(t)$ are
marginal and, besides the standard factor ${1\over 1-z}$, the main
$z$ dependence of the flux comes from the  kinematical boundary
$|t| \geq |t|_{min}= {m_{p}^{2}
(1-z)^{2}\over z}$ in the form factor  $G(t)$.
In principle $F_{2,\Pom}^{D(4)}(z,t,\beta,Q^2)$ can
be derived from experimental data on diffractive DIS, but
currently for $\beta \lsim 2\cdot 10^{-3}-10^{-2}$ the
pomeron structure function stays basically unknown.
It has been argued, \cite{NZ94}, that at small $\beta$ the conventional DGLAP
evolution holds for the pomeron structure function giving a
  $\beta,$ and $Q^{2}$ dependence
of $F_{2}^{\Pom}(\beta,Q^{2})$  similar
to that of $F_{2}^{\pi}(\beta,Q^{2})$ (see for instance \cite{GNZ95}).
On the other hand, the triple-pomeron formula with soft pomerons
gives the scaling prediction $F_{2}^{\Pom}(\beta,Q^{2}) =
C_{\Pom} \beta^{-0.08}$
(the normalization $C_{\Pom}=0.026$ has been fitted \cite{GKS97}
to the H1 experimental data \cite{H1_reggeon97}). We use these two models to
check the sensitivity of the leading proton spectra to the evolution in
$\beta$ and $Q^{2}$.

The reggeon exchange is an important ingredient of the triple-Regge
phenomenology of hadronic diffraction, although its strength
is not very well known\cite{Kaidalov,FFox,KKLP76}. The triple-Regge
parameterization for the reggeon flux is
\begin{equation}
f_{R}(z,t) = \frac{1}{8 \pi^{2}} (1-z)^{1-2\alpha_R(t)} G_{R}(t)\,
\; ,
\end{equation}
where $\alpha_R(t)$ is the reggeon trajectory. We take
$G_{R}(t) = G_{R}(0) \cdot \exp(B_{R}t)$, where for the dominant
$f_{0}$-exchange $G_{R}(0) = 76$ mb \cite{NSZ96,GKS97} and
$B_{R}$ = 4 $\,$GeV$^{-2}$, which is consistent with the data on leading
proton production in $pp$ collisions \cite{Barton}. Triple-Regge
considerations in conjunction with
fits to the NN total cross sections \cite{DLsigma} suggest
the isovector reggeon exchange to be much weaker than the isoscalar
$f_{0}$ exchange \cite{NSZ96,GKS97}.
The reggeon structure function is basically unknown. The extension
of microscopic analysis \cite{NZ94} to reggeons suggests that
the $\beta$ and $Q^{2}$ dependence of this structure function at small
$\beta$ must be similar to that of the pion structure function,
$F_{2}^{R}(\beta,Q^{2}) \sim F_{2}^{\pi}(\beta,Q^{2})$ and/or the
pomeron structure function, $F_{2}^{R}(\beta,Q^{2}) \sim
F_{2}^{\Pom}(\beta,Q^{2})$. Arguably the gross features of the
small-$\beta$, large-$Q^{2}$ behavior of $F_{2}^{i}(\beta,Q^{2})$
should be similar for all targets $i$. If the
extrapolation along the Regge trajectory to the particle pole
$t=m_{f_{0}}^{2}$ were possible, one could have related
$F_{2}^{R}(\beta,Q^2)$
to the $f_{0}$ meson structure function, which at small
$\beta$ is expected to
be similar to $F_{2}^{\pi}(\beta,Q^{2})$.
Going from the particle pole $t = m_{f_{0}}^{2}$ to the
scattering region $t<0$ brings the off-mass shell suppression in, and it is
natural to expect $F_{2}^{R}(\beta,Q^{2}) < F_{2}^{\pi}(\beta,Q^{2})$. The
triple-Regge phenomenology of hadronic diffraction suggests the suppression
factor $\lambda_{f} \sim 0.5$ with a large uncertainty \cite{NSZ96}
(because of different normalization of the flux
in \cite{NSZ96} and \cite{GKS97},
the estimate of $\lambda_{f}$ in \cite{NSZ96} must be taken with the
factor ${\pi \over 2}$). Similarly to the pomeron structure function,
the triple-Regge formalism with soft pomerons gives the scaling
$F_{2}^{R}(\beta,Q^{2}) = C_{R}\beta^{-0.08}$. Eventually,
with high precision data on diffractive DIS, one would be able to evaluate
$F_{2}^{R}(\beta,Q^{2})$ directly from the reggeon background to
pomeron exchange at $z \gsim 0.95$.

The single particle inclusive $(z,t)$-spectrum of protons is defined as
$R(z,t,x,Q^{2}) = F^{(4)}(z,t,\beta,Q^{2})/F_{2p}(x,Q^{2})$.
A fully differential study of $R(z,t,x,Q^{2})$ is not yet possible with
the limited statistics of the preliminary ZEUS data \cite{ZEUS_LPS97}.
The data were collected  within the following experimental cuts $\Omega_{exp}$:
$0.6 < z < 0.9$, $|t|_{min} < |t| < 0.5 $GeV$^{2}$, $ 10^{-4} < x < 10^{-3}$
and $4 < Q^{2} < Q_{max}^{2}$, where $Q_{max}^{2}$ is the maximal
kinematically attainable $Q^{2}$.
Within these cuts  the fraction of events with leading proton is given by:
\begin{equation}
R_{exp} = \sum_{i} R_{exp}^{i} =
\sum_{i}\frac{\Delta \sigma^{i}(\Omega_{exp})}
{\Delta \sigma^{tot}(\Omega_{exp})} \; ,
\label{Ratio}
\end{equation}
where
\begin{equation}
\Delta \sigma^{i}(\Omega_{exp}) =
\int_{z_{min}}^{z_{max}}  dz   \int_{t_{min}}^{t_{max}} dt
\int_{x_{min}}^{x_{max}}  dx   \int_{Q_{min}^{2}}^{Q_{max}^{2}}  dQ^2
\frac{d \sigma^{i}}{dx dQ^2 dz dt}\, ,
\label{Sigma_LP}
\end{equation}
\begin{equation}
\Delta \sigma^{tot}(\Omega_{exp}) =
\int_{x_{min}}^{x_{max}}  dx  \int_{Q_{min}^{2}}^{Q_{max}^{2}}  dQ^2
\frac{d \sigma^{tot}}{dx dQ^2} \; ,
\label{Sigma_inc}
\end{equation}
and the subscript $i$ stands for one of the mechanisms shown in Fig.1.

As emphasized above, the pion, pomeron and reggeon structure
functions are unknown in the  $\beta$ region considered  in our present
analysis. For a reference evaluation of $R_{exp}^{i}$, we take the
GRV parameterization for the ($\beta,Q^{2}$) evolution of the pion structure
function \cite{GRV_pion}, the flux of pions evaluated in the light-cone
model for the chiral structure of the nucleon \cite{HSS96}
(the Regge parameterization leads to very similar result \cite{NSZ97})
and the triple-Regge model parameterization $F_{2}^{\Pom}(\beta,Q^{2})=
0.026\beta^{-0.08}$ described above. For the proton
structure function, which enters the evaluation of the denominator
(\ref{Sigma_inc}) of the ratio (\ref{Ratio}), one can use any
convenient fit to the HERA data. In the present analysis we take the GRV
parameterization \cite{GRV_proton}. The practical calculations have been
performed with a Monte Carlo implementation of the above formalism.
As a result of our analysis, we find $R_{\pi^{0} p}(\Omega_{exp})$ = 2\%,
$R_{\pi \Delta}(\Omega_{exp})$ = 0.9\%
and the tail of the pomeron exchange contribution \cite{GK96} gives
$R_{\Pom p}(\Omega_{exp})$ = 1.2\%, so that  $R_{1+2+3}(\Omega_{exp})$
 = 4.05\%. Note that QCD hadronization models (Ariadne, Herwig) were never
meant to describe the nonperturbative proton
fragmentation region; for example
Ariadne \cite{Ariadne} gives $R(\Omega_{exp})$
in the per mill range and a similar under-prediction for the
production of leading neutrons \cite{ZEUSneutron}.
>From the comparison with the ZEUS experimental result,
$R^{ZEUS}_{exp}$ = 9.2 $\pm$ 1.7 \%(stat. only) \cite{ZEUS_LPS97},
we conclude that about
5 \% of the missing strength must be attributed to the reggeon
exchange. In the triple-Regge scaling model,
$F_{2}^{R}(\beta,Q^{2})=C_{R}\beta^{-0.08}$,
this requires $C_{R} = 0.12$ within a factor of 1.5 uncertainty.

The importance of different mechanisms can be better seen from
the z-dependence of the ratio $R_{exp}(z)$ defined for the experimental
$(t,x,Q^{2})$ range as shown in Fig.~2. Clearly the importance
of the reggeon exchange can be seen from the figure.
With the set of parameters specified above, the reggeon contribution
makes $R_{exp}(z)$ approximately flat at $z\lsim 0.9$, in close similarity
to a flat $z$-spectrum of leading protons in hadronic interactions
\cite{Barton}. The preliminary H1 results are also consistent with the flat
$z$-spectrum \cite{H1_LPS97}.

The $z$-spectrum of  leading nucleons from
diffraction double dissociation (DD)
has been studied in \cite{HNSSZ96}. It can be isolated experimentally
by the rapidity gap (GAPCUT) selection method \cite{ZEUS_LPS97}. An extension
of the analysis \cite{HNSSZ96} to leading protons shows that $\sim 70\%$
of DD  events have leading protons, mainly
produced by  excitations of the $N\pi\pi$ and
high mass continuum states. Roughly  $\sim 50 \%$ of final state protons
have $z>0.6$. Since DD constitutes
$\sim 2\%$ of the  DIS events, only $\sim 0.7\%$ have a leading proton with
$z>0.6$ generated by this mechanism. DD is therefore a small,
$f({\rm GAPCUT})=R_{DD}(\Omega_{exp})/R_{exp}
\sim 7\%$, background to the dominant non-diffractive
production mechanism, in good agreement with the ZEUS findings.
The LEPTO6.5 `soft
color interaction' model \cite{LEPTO} which, unlike Ariadne and/or Herwig,
is supposed to describe all aspects of DIS including leading proton
production, over-predicts the fraction of GAPCUT events: $f$(GAPCUT)$
\approx$20-30\%. The observed leading proton $z$-distribution
of the GAPCUT sample \cite{ZEUS_LPS97}  is also consistent
with the z spectrum of protons generated in proton dissociation
into $N\pi\pi$ and continuum states as shown in
Fig.~2 of \cite{HNSSZ96}. It can readily be included in the analysis
of higher precision data.

This evaluation of the reggeon exchange from fragmentation into protons
at $z \lsim 0.9$ is consistent within a factor  of 2 with estimates
of the reggeon background to pomeron exchange in the diffractive region
of $z \gsim 0.95$ \cite{GKS97}. A caveat in comparing
these two extreme regions is
the possible reggeon-pomeron interference contribution
$\sim {1\over \sqrt{1-z}}$, which can be substantial in the diffractive
domain and small in the fragmentation region $z \lsim 0.9$.
A combined analysis of high precision fragmentation and
diffractive data would be
the best way to fix the reggeon-pomeron interference
contribution, but such an
involved phenomenology is not warranted with the
presently available data. Note also that the ZEUS
data are preliminary and lacking the evaluation of systematic errors.

In Fig.~3 we show the slope $b(z)$ of the $t$-distributions defined
in the experimental range of $(x,Q^{2})$ ($R(z,t) \propto exp(b(z)t)$).
The slope of the reggeon trajectory is large, $\alpha_{R}' = 0.9$ GeV$^{-2}$,
and for pure reggeon exchange contribution quite a substantial rise of
the slope is expected: $b_{R}(z) = B_{R} + 2\alpha_{R}' \log{1\over 1-z}$.
Similar growth of the slope is expected also for the pion exchange
contribution. The increase of the slope at large $z$ is tamed by the small
diffraction slope of the pomeron contribution.
The parameter $B_{R}$ is poorly known and the leading proton spectrum offers
the best possibility for its determination. Our results
for $b(z)$ obtained with $B_{R}=4$ GeV$^{-2}$ are close to the slope
of the $t$-dependence for leading protons observed in $pp$ collisions
\cite{Barton}.

The ($x,Q^{2}$)-dependence of different mechanisms is controlled by
the ratios $\rho_{i}(x,Q^{2})=F_{2}^{i}({x \over 1-z},Q^{2})/F_{2p}(x,Q^{2})$.
In the scenario with the scaling soft pomeron/reggeon
structure functions, $\rho_{\Pom,R}(x,Q^{2})$
decreases  with rising $Q^{2}$ and/or decreasing $x$, because of
the scaling violations and steep $x$-dependence of the proton
structure function $F_{2p}(x,Q^{2})$.
Fig.~4 shows that in such a scaling scenario
(no QCD evolution for $F_{2}^{\Pom,R}(\beta,Q^{2}$)),
one would expect significant dependence of the
leading proton production on both $x$ and $Q^2$.
On the other hand, if $F_{\Pom,R}(\beta,Q^{2})$ satisfies the conventional
$(\beta,Q^{2})$ evolution at small $\beta$ \cite{NZ94}, one would expect very
weak $(x,Q^{2})$ dependence of the leading proton spectrum. This stays true
also in the real photoproduction limit. In the present analysis we model
the evolution effects by taking
$F_{2}^{\Pom,R}(\beta,Q^{2}) = \lambda_{\Pom,R} f_{2}^{\pi}(\beta,Q^{2})$
with the GRV pion structure function.
We adjust $\lambda_{\Pom} = 0.2$ and $\lambda_{f} = 0.5$ as it was
evaluated in \cite{NSZ96}
so that we reproduce the same $R_{exp}^{i}$ as with the scaling (no evolution)
scenario within the ZEUS kinematical cuts.

In the conventional evolution scenario we indeed find a very weak
$x$ and $Q^{2}$ dependence of the leading proton spectra. The preliminary
ZEUS data \cite{ZEUS_LPS97} better agree with this scenario, although the
error bars are still rather large. The preliminary H1 data \cite{H1_LPS97}
on the $t$-integrated $F_{2}^{(4)}(z,t,x,Q^{2})$ also support the
conventional evolution scenario.


We conclude that the salient features of fragmentation into leading
protons can be understood quantitatively in terms of peripheral
mechanisms  extended to the DIS regime.
The experimentally observed similarity of the leading proton spectra
in  $pp$ collisions and $ep$ DIS is a natural consequence of these
mechanisms. We emphasize
that our approach has the capability of a unified description of
diffractive DIS at $z \gsim 0.9$ and of fragmentation into
protons in non-diffractive DIS. Of the four sources of leading
protons pion exchange  can be experimentally determined
using neutron tagged DIS. Experimental
confirmation of our estimate for this process
will lend strong support also for our evaluation of the $\Delta$
contribution. The pomeron exchange background
can be inferred from diffractive DIS. Finally,
the reggeon exchange mechanism of fragmentation can also be tested
in diffractive DIS. The combined analysis of the high precision
leading proton data and diffractive DIS data makes possible
a determination of the reggeon-pomeron interference effects,
which has not been accomplished with the hadronic diffraction data
\cite{Kaidalov,FFox,KKLP76}.

A comparison of the soft pomeron no-evolution (unrealistic though it
is) and conventional evolution scenarios for the reggeon structure
function shows that the high precision leading proton spectrum
offers an interesting test of the universality of the QCD evolution
properties of structure functions at small $x$ (For a related
discussion of the pion exchange mechanism within Veneziano's fracture
function
\cite{Veneziano} context see \cite{Florian}.).

We conclude with the comment that similar fragmentation mechanisms
may be at work also at smaller $z$, where DIS on the multi-pion
Fock states of the physical nucleon, $(n\pi)N,(n\pi)\Delta,(n\pi)N^{*},
(n\pi)\Delta^{*}$, provides a natural mechanism for slowing down
secondary protons. In the spirit of the above discussion, the weak $x,Q^{2}$
dependence must hold also for slower protons. Similar arguments hold
for the fragmentation of protons into hyperons ($\Lambda,\Sigma,$...).

\vskip 1cm

\noindent
{\bf Acknowledgments}
At all stages of this work we benefited much from the continuous 
advice of, and enlighetning discussion with, N.Cartiglia of the ZEUS 
collaboration.
One of us (A.S.) is indebted to K. Golec-Biernat for interesting discussion.
This work was supported partly by the German-Polish exchange program,
grant No. POL-81-97 and by the INTAS Grant 97-0597.




\newpage

{\large \bf Figure captions}

Fig.1\\
Peripheral mechanisms of leading proton production.\\

\noindent
Fig.2\\
The fraction (in per cent) of DIS events with a leading proton in a given
$z$ bin ($\Delta z =0.03$) predicted by our model (thick solid curve) in
comparison with the ZEUS preliminary data \cite{ZEUS_LPS97}. The
contributions of four mechanisms of Fig.~1 are shown separately:
the thin solid line shows the pomeron-exchange contribution,
the long-dashed curve is for the pion-exchange contribution, the
dashed curve shows protons from the $\Delta$ production and the
dotted curve is for the reggeon-exchange component.\\

\noindent
Fig.3\\
The slope of the t-distributions predicted by the model is
compared with the ZEUS preliminary data \cite{ZEUS_LPS97}.\\

\noindent
Fig.4a\\
The sensitivity of the fraction of DIS events containing leading protons
within ZEUS cuts to the $Q^{2}$ evolution effects for the two different
scenarios for the pomeron and reggeon structure function: the curves marked
by filled circles are for the no-evolution soft pomeron model,
the unmarked curves show the results for the conventional QCD
evolution scenario modeled by the GRV pion structure function.
The legend of curves is the same as in Fig.~1.\\

\noindent
Fig.4b\\
The same as Fig.~4a, but for the $x$-dependence of the fraction of
DIS events containing leading protons within ZEUS cuts.


\begin{thebibliography}{99}

\bibitem{BEBC}
BEBC Collaboration. J. Guy et al., Phys. Lett. {\bf B229} (1989) 421.

\bibitem{MNT}
W. Melnitchouk, N.N. Nikolaev and A.W. Thomas, Z. Phys.
{\bf A342} (1992) 215;\\
G.D. Bosveld, A.E.L. Dieperink and A.G. Tenner, Phys. Rev.
{\bf C49} (1994) 2379;\\
A. Szczurek, G.D. Bosveld and A.E.L. Dieperink, Nucl. Phys.
{\bf A595} (1995) 307.

\bibitem{ZEUS_LPS97}
N. Cartiglia for the ZEUS collaboration, a talk at DIS97, Chicago
( April 1997), Proceedings, American Institute of Physics,
editors D.Krakauer and J.Repond, in press, hep-ph/9706416;\\
ZEUS Collaboration, a talk (N-644) at the International Europhysics Conference
on High Energy Physics, Jerusalem, 19-26 August 1997.

\bibitem{H1_LPS97}
H1 Collaboration, a talk (abstract 379) at the International Europhysics
Conference on High Energy Physics, Jerusalem, 19-26 August 1997.

\bibitem{Ariadne}
L. L\"onblad, Comput. Phys. Commun. {\bf 71} (1992) 15;
B. Andersson et al., Phys. Rep. {\bf 97} (1983) 31.

\bibitem{Herwig}
G. Marchesini et al., Comput. Phys. Commun. {\bf 67} (1992) 465.

\bibitem{DeWolf}
E.A. DeWolf, A.T. Doyle, N. Varelas and D. Zeppenfeld, Summary of the
Hadronic Finals States Working Group at the DIS97 Workshop,
Chicago (April 1997),

\bibitem{ZEUSneutron}
ZEUS Collaboration; M. Derrick et al.
Phys. Lett. {\bf  B 384} (1996) 388.

\bibitem{Kaidalov}
A.B. Kaidalov, Phys. Reports {\bf 50} (1979) 157.

\bibitem{NSZ97}
N.N. Nikolaev, J. Speth and B.G. Zakharov. KFA-IKP-TH-1997-17,
hep-ph/9708290.

\bibitem{ST}
J. Speth and A.W. Thomas, KFA Report J\"ul-3283 (1996), Adv. Nucl. Phys.
(1997) in press.

\bibitem{HSS96}
H. Holtmann, A. Szczurek and J. Speth, Nucl.Phys.{\bf A596} (1996) 631.

\bibitem{DHSS97}
Z. Dziembowski, H. Holtmann, A. Szczurek and J. Speth,
Ann. Phys. (N.Y.) {\bf 258}
(1997) 1.

\bibitem{MC_he}
A. Szczurek, V. Uleshchenko, H. Holtmann and J. Speth,
Nucl. Phys.{\bf A624} (1997) 495;\\
A. Szczurek and A. Budzanowski, Phys. Lett.{\bf B404} (1997) 141;\\
A. Szczurek and A. Budzanowski, Phys.Lett.{\bf B408} (1997) 275.

\bibitem{SEHS96}
H. Holtmann, N.N. Nikolaev, J. Speth and A. Szczurek, Z. Phys.
{\bf A353}, 411 (1996);
A. Szczurek, M. Ericson, H. Holtmann and J. Speth, Nucl. Phys.{\bf A596}
(1996) 397.

\bibitem{HLNSS94}
H. Holtmann, G. Levman, N.N. Nikolaev, A. Szczurek and J. Speth,
Phys.Lett.{\bf B338} (1994) 363.

\bibitem{Lusignoli}
M. Lusignoli, Nucl. Phys. {\bf B138} (1978) 151.

\bibitem{KPP96}
B. Kopeliovich, B. Povh and I. Potashnikova, Z. Phys. {\bf C73} (1996) 125.

\bibitem{PSI97}
M. Przybycie\'n, A. Szczurek and G. Ingelman, Z. Phys. {\bf C74} (1997) 509.

\bibitem{H1_reggeon97}
H1 Collaboration, T. Adloff et al., preprint DESY97-158,
submitted to Z. Phys.C

\bibitem{NSZ96}
N.N. Nikolaev, W. Sch\"afer and B.G. Zakharov, hep-ph/9608338, unpublished.

\bibitem{GK96}
K.Golec-Biernat and J.Kwieci\'nski, Phys.Rev. {\bf D55} (1997) 3209.

\bibitem{GKS97}
K.Golec-Biernat, J.Kwieci\'nski and A.Szczurek, Phys. Rev. {\bf D56} (1997)
3995.

\bibitem{HNSSZ96}
H. Holtmann, N.N. Nikolaev, A. Szczurek, J. Speth and B.G. Zakharov,
Z. Phys. {\bf C69}, 297 (1996).

\bibitem{ZakSerg}
B.G. Zakharov and V.N. Sergeev, Yad. Fiz. {\bf 39} (1984) 707;
G.R. Goldstein and J.F. Owens, Nucl. Phys. {\bf B118} (1977) 29.

\bibitem{NZ94}
N.N.Nikolaev and B.G.Zakharov, Z. Phys. {\bf C64} (1994) 631.

\bibitem{GNZ95}
M. Genovese, N.N. Nikolaev and B.G. Zakharov, J. Exp. Theor. Phys.
{\bf 81} (1995) 625; {\bf 81} (1995) 633.

\bibitem{DLsigma}
A. Donnachie and P.V. Landshoff, Phys. Lett. {\bf B296} (1992) 227.

\bibitem{FFox}
R.D. Field and G.C. Fox, Nucl. Phys. {\bf B80} (1974) 367.

\bibitem{KKLP76}
Yu.M. Kazarinov, B.Z. Kopeliovich, L.I. Lapidus and I.K. Potashnikova,
Sov. Phys. JETP, {\bf 43} (1976) 598.

\bibitem{DLdif}
A. Donnachie and P.V. Landshoff, Phys. Lett. {\bf B199} (1987) 309.

\bibitem{Barton}
A.E. Brenner et al., Phys. Rev. {\bf D26}, 1497 (1982)

\bibitem{GRV_pion}
M. Gl\"uck, E. Reya, A. Vogt, Z. Phys. {\bf C53} (1992) 651.

\bibitem{GRV_proton}
M. Gl\"uck, E. Reya, A. Vogt, Z. Phys. {\bf C67} (1995) 433.

\bibitem{LEPTO}
G. Ingelman et al., Comput. Phys. Commun. {\bf 101} (1997) 108.

\bibitem{Veneziano}
L. Trentadue and G. Veneziano, Phys. Lett. {\bf B323} (1994) 201;

\bibitem{Florian}
D. De Florian and R. Sassot, To be published in the proceedings of
Madrid Workshop on Low x Physics, Madrid, Spain, 18-21 Jun 1997,
hep-ph/9710205.

\end{thebibliography}
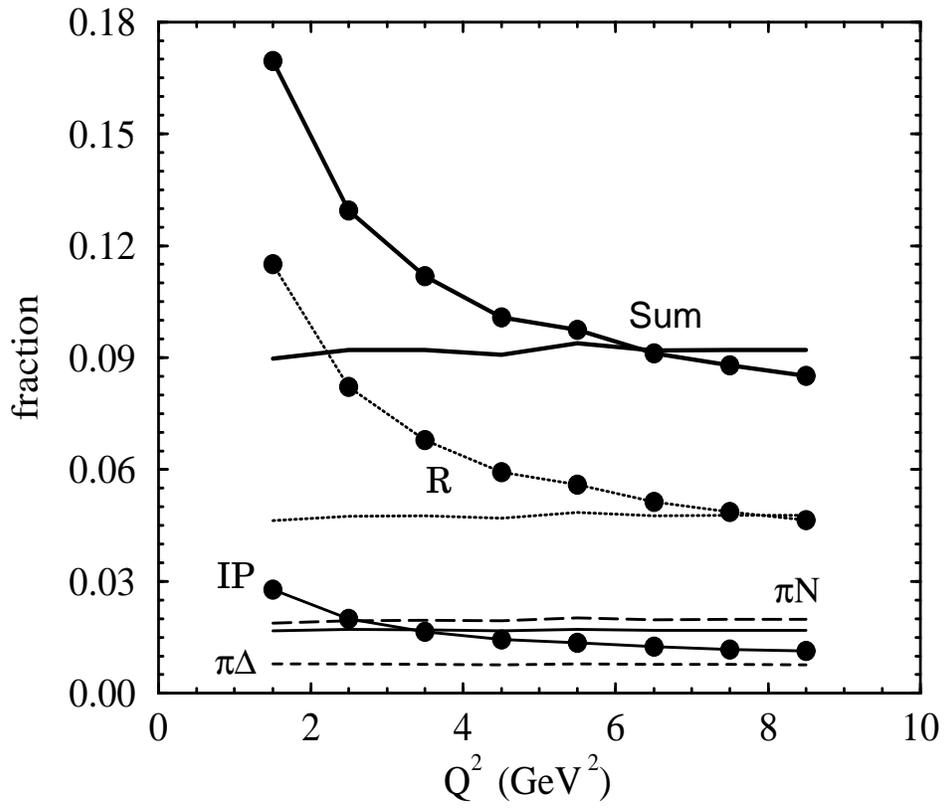
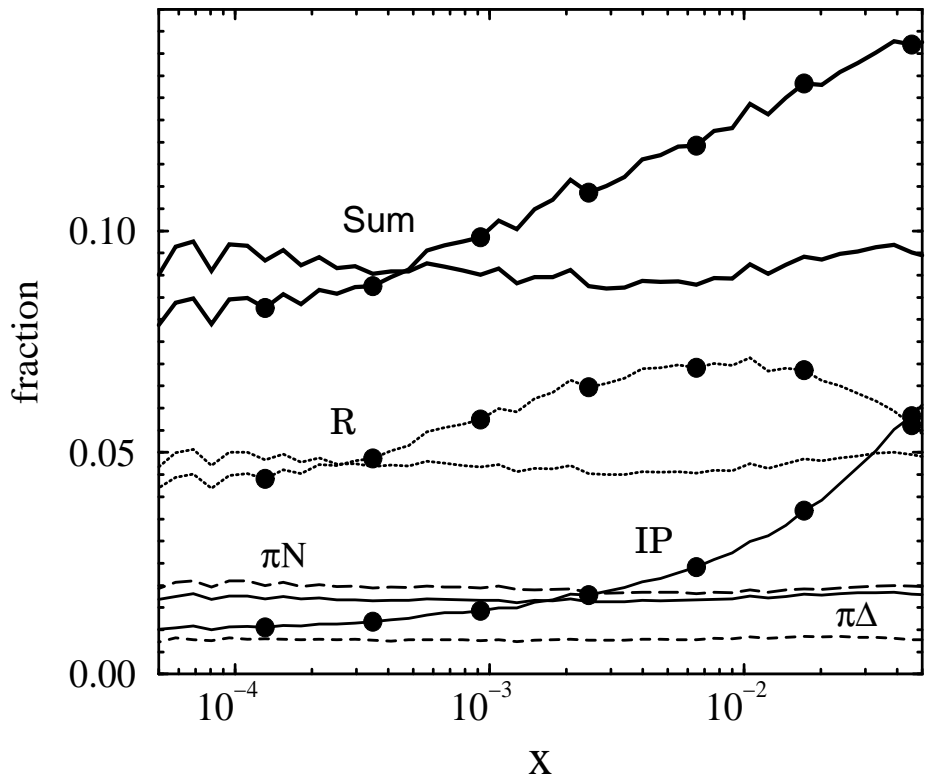
\end{document}